# Computation of spontaneous emission dynamics in colored vacua

M.H. Teimourpour and R. El-Ganainy[*]

*Department of Physics, Michigan Technological University, Houghton, Michigan 49931, USA*

**Abstract**

We present an efficient time domain numerical scheme for computing spontaneous emission dynamics in colored vacua. Starting from first principles, we map the unitary evolution of a dressed two-level quantum emitter onto the problem of electromagnetic radiation from a complex harmonic oscillator under self-interaction conditions. This latter oscillator-field system can be efficiently simulated by using finite difference time domain method without the need for calculating the photonic eigenmodes of the surrounding environment. In contrast to earlier investigations, our computational framework provides a unified numerical treatment for both weak and strong coupling regimes alike. We illustrate the versatility of our scheme by considering several different examples.

PACS#  02.60.Cb, 42.50.Pq ,  42.50.Ct

[*] ganainy@mtu.edu

Spontaneous emission (SE) is a fundamental process in quantum physics that plays a major role in several applications [1, 2]. In the early days of quantum mechanics, spontaneous emission was believed to be an intrinsic property of the atom. However, the seminal work by Purcell [3] showed that this process is governed by the electromagnetic (EM) local density of states (LDOS) of the surrounding medium. This work has inspired many subsequent theoretical studies [4-6] and experimental investigations [7, 8]. In this regard, the recent unprecedented progress in fabrication and measurement technologies has provided a means for controlling spontaneous emission of quantum dots in both weak and strong coupling regimes [9]. The former is characterized by a small perturbation over the vacuum photonic local density of states which leads to a modification of the exponential radiative decay rate- a process that is known as Purcell effect. The latter, on the other hand, is marked by LDOS singularities associated with colored vacua such as photonic band edges [10] or isolated discrete modes [11]. While several methods exist to calculate the Purcell effect [12, 13], treating the strong coupling regime can be a challenging task that requires a full knowledge of the photonic eigenmodes of the system and their optical dispersion. Noteworthy, for a quantum dot located inside high quality optical resonators, light-matter interaction can be approximated by an effective non-Hermitian Jaynes-Cumming (JC) Hamiltonian that can be solved analytically [14]. However, to do so, both the optical eigenmode and the cavity quality factor have to be determined first. Moreover, treating the boundary between the weak and strong coupling regimes is not as straightforward. This is due to the proximity of this intermediate domain to special algebraic singularities known as the exceptional points (EPs) [15] at which Fermi's Golden rule was shown to break down [16]. This can occur for instance if the quantum emitter (QE) interacts strongly with large number of optical modes near photonic band edges [10]. In order to overcome some of these difficulties, approaches based on quasi-normal modes (QNM) and constant flux (CF) states were proposed [17, 18]. However, numerical evaluation of these leaky eigenmodes for any arbitrary photonic structure can be complicated [19].

In this letter we present a new approach for investigating the problem of spontaneous emission of a single emitter located inside (or in the vicinity of) any arbitrary photonic structure. Our scheme is general and treats all regimes of operation on equal foot. Most importantly, it does not require any knowledge of the real Hermitian eigenmodes of the systems nor of its non-Hermitian QNM's or CF states. In particular, we start by demonstrating that the unitary evolution of a dressed two-level quantum emitter can be mapped onto a complex radiative harmonic oscillator under self-interaction conditions. We note of course that treating SE by using classical Hertzian dipoles dates back to early days of quantum mechanics [20]. However, despite dealing with small number of photons, these previous investigations involved some heuristic mean field approximations for the optical fields as noted for instance in Ref. [21]: "*It should not be overlooked that the use of the classical field relation and the introduction of an unspecified damping term at this point is a simplifying assumption for escaping the difficulties of a quantum field theoretical approach*". What distinguishes our work here is that we carry out our analysis without making any such assumptions.

To this end, we consider the Hamiltonian: $\hat{H} = \hat{H}_A + \hat{H}_R + \hat{H}_I$, where $\hat{H}_{A,R,I}$ are the atomic, radiation and interaction parts, respectively. The radiation Hamiltonian depends only on the photonic environment and is given by $\hat{H}_R = \sum_k \hbar\omega_k \hat{a}_k^+ \hat{a}_k$, where $\omega_k$ is the frequency, $\hat{a}_k^+$ and $\hat{a}_k$ are creation and annihilation operators associate with an optical mode characterize by a set of parameters denoted by $k$ and as usual $\hbar$ is Planck's constant. For a two-level system, shown schematically in Fig.1 (a), the atomic Hamiltonians takes the form $\hat{H}_A = \hbar\omega_o |e\rangle\langle e|$ where $\omega_o$ is the atomic transition frequency between the excited state $|e\rangle$ and the ground state $|g\rangle$ whose energy is taken to be zero. Within the dipole approximation, the interaction Hamiltonian between the quantum emitter and the EM fields reads $\hat{H}_I = \sum_k \hbar(\hat{a}_k^+ + \hat{a}_k)(\gamma_k |g\rangle\langle e| + \gamma_k^* |e\rangle\langle g|)$, where

$\gamma_k = -\sqrt{\dfrac{\omega_k}{2\varepsilon_0 \hbar V}} \vec{E}_k(\vec{r}_o) \cdot \vec{d}_{eg}$ quantifies the strength of radiation-matter coupling and the asterisks superscript denote complex conjugate. In the above, $\vec{E}_k(\vec{r})$ is the normalized electric field distribution associated with mode $k$ and is obtained by solving Maxwell's equations under the appropriate boundary conditions [22] whereas $\vec{r}_o$ is the location of the quantum emitter and $\vec{d}_{eg}$ is electric-dipole transition matrix element between the two levels. Note that we have not made the rotating wave approximation in $\hat{H}_I$. In order to construct an approximate model of this two-level system based on classical radiative oscillators, we consider the spontaneous emission from an equidistant three-level system when the electron is initially in the first excited state as shown in Fig.1 (b). Note that because all EM radiation modes are initially in the vacuum state, the evolution dynamics of these two different systems are nearly identical. Under these conditions, the new atomic and interaction Hamiltonians can be written as: $\hat{H}_A = \hbar\omega_o |e_1\rangle\langle e_1| + 2\hbar\omega_o |e_2\rangle\langle e_2|$ and $\hat{H}_I = \sum_k \hbar(a_k^+ + a_k)\{\gamma_k|g\rangle\langle e_1| + \tilde{\gamma}_k|g\rangle\langle e_2| + \gamma_k|e_1\rangle\langle e_2| + c.c\}$, where the states $|e_1\rangle$ and $|e_2\rangle$ are the first and second excited states, correspondingly. Note that in this artificial model we assumed that the photonic mediated coupling strength between the states $|g\rangle$ and $|e_1\rangle$ is identical to that between $|e_1\rangle$ and $|e_2\rangle$. On the other hand, we have not made any assumption about the value of $\tilde{\gamma}_k$ since it does not play any role on our analysis. Throughout the rest of this work, we will refer to the above three-level Hamiltonians.

Next we investigate the unitary evolution of this system shown in Fig.1 (b) under the initial condition $|\psi(t=0)\rangle = |e_1,0\rangle$ that represents an atom in the first excited state with no photons. We do so by using a variational wavefunction expansion: $|\psi(t)\rangle = a(t)|e_1,0\rangle + \sum_k b_k(t)|g,1_k\rangle + \sum_k c_k(t)|e_2,1_k\rangle$, where $a(t)$, $b(t)$ and $c(t)$ are the projection coefficients on the corresponding states. Note that the second term in the expansion is responsible for the SE decay rate while the third term represents a non-resonant contribution. By substituting $|\psi(t)\rangle$ in Schrodinger's equation, we obtain:

$$i\frac{da(t)}{dt} = \omega_o a(t) + \sum_k \gamma_k b_k(t) + \sum_k \gamma_k c_k(t)$$
$$i\frac{db_k(t)}{dt} = \omega_k b_k(t) + \gamma_k^* a(t) \qquad (1)$$
$$i\frac{dc_k(t)}{dt} = (2\omega_o + \omega_k)c_k(t) + \gamma_k^* a(t)$$

The Laplace transform of Eq. (1) is:

$$isA(s) - ia(0) = \omega_o A(s) + \sum_k \gamma_k B_k(s) + \sum_k \gamma_k C_k(s)$$
$$isB_k(s) = \omega_k B_k(s) + \gamma_k^* A(s) \qquad (2)$$
$$isC(s) = (2\omega_o + \omega_k)C_k(s) + \gamma_k^* A(s)$$

where $A(s)$, $B(s)$ and $C(s)$ are the Laplace transforms of $a(t)$, $b(t)$ and $c(t)$, correspondingly. Note that $b(0) = c(0) = 0$. By eliminating $B_k(s)$ and $C_k(s)$ from Eq.(3) we arrive at $isA(s) - \omega_o A(s) + \Gamma_s(s)A(s) = ia(0)$, where the self-interaction term is given by $\Gamma_s(s) = \sum_k \left( \frac{|\gamma_k|^2}{is - \omega_k} + \frac{|\gamma_k|^2}{is - 2\omega_o - \omega_k} \right)$. In principle, one can obtain the SE dynamics by taking the inverse Laplace transform of $A(s)$. However, in the most general case this is not a straightforward task. In addition, in order to calculate $\Gamma_s(s)$, all the optical eigenmodes and their dispersion relations must be first computed. In order to circumvent these difficulties, we now construct an equivalent system of equations that can be numerically solved by standard time-domain EM solvers.

Conceptually, we start by assuming that a quantum emitter can be represented by some form of a radiative oscillator coupled to electromagnetic radiation. In particular, we consider Maxwell's equations of the following form:

$$\mu_0 \frac{\partial \vec{H}}{\partial t} = -\nabla \times \vec{E}$$

$$\varepsilon_o \varepsilon_r(\vec{r}) \frac{\partial \vec{E}}{\partial t} = \nabla \times \vec{H} - N_D \delta_D(\vec{r} - \vec{r}_o) \int_0^t \vec{f}(t')dt' \qquad (3)$$

$$i \frac{d\vec{f}(t)}{dt} - \omega_o \vec{f}(t) = -\frac{d_{eg}^2}{\hbar} \omega_o^2 \vec{E}(\vec{r}_o, t)$$

where $\mu_o$ and $\varepsilon_o$ are the free space permeability and permittivity, respectively and $\varepsilon_r(\vec{r})$ is the position dependent dielectric function. Here, $\vec{f}(t)$ represents the time variation of the source and its oscillation direction. In addition in Eq (3), the subscript $D$ indicates the dimensionality of the problem (and consequently the delta function) and the normalization factor $N_D$ takes the following values: $N_1 = 1/A$, $N_2 = 1/L$ and $N_3 = 1$, where $L$ and $A$ are the axial length and cross section area of 2D and 1D quantum emitters as shown schematically for some realistic structures in Fig.2.

By taking the Laplace transform of Eq.(3) and solve for $\vec{F}(S) = L\{\vec{f}(t)\}$ we find that

$$is\vec{F}(s) - \omega_o \vec{F}(s) + \tilde{\Gamma}_s(s)\vec{F}(s) = i\vec{f}(0), \text{ where } \tilde{\Gamma}_s(s) = \sum_k \left( \frac{|\gamma_k|^2}{is - \omega_k} - \frac{|\gamma_k|^2}{is + \omega_k} \right) \text{ and we have}$$

dropped the reference to the oscillator location $\vec{r}_o$. In arriving at these results, we have made use the Laplace transform of the retarded EM Green's function [22]. Note that the most significant contribution to the resonant terms of $\Gamma_s(s)$ and $\tilde{\Gamma}_s(s)$ occurs around the point $is \approx \omega_o$ where also the nonresonant terms becomes approximately identical. Consequently, spontaneous emission can be studied by using the oscillator-field system of Eq.(3) instead of the original quantum problem described by Eq.(1). Additionally, Eq.(3) can be further simplified by substituting $\vec{f}(t) = \frac{d^2}{dt^2}\vec{P}(t)$ and noting that the time dynamics of $\vec{P}(t)$ includes two different time scales: a rapid component that oscillates at a frequency $\omega_o$ and a relatively slower decay rate. A factorization of the rapid variations in the oscillator equation through the substitution $\frac{d^2}{dt^2} \approx -\omega_o^2$ gives:

$$\mu_0 \frac{\partial \vec{H}}{\partial t} = -\nabla \times \vec{E}$$

$$\varepsilon_o \varepsilon_r(\vec{r}) \frac{\partial \vec{E}}{\partial t} = \nabla \times \vec{H} - N_D \delta_D(\vec{r} - \vec{r}_o) \frac{d\vec{P}(t)}{dt} \quad (4)$$

$$i \frac{d\vec{P}(t)}{dt} - \omega_o \vec{P}(t) = \frac{d_{eg}^2}{\hbar} \vec{E}(\vec{r}_o, t)$$

Equation (4) now takes a more familiar form where $\vec{P}(t)$ represents the dipole moment associated with the quantum emitter and the quantity $N_D \delta_D(\vec{r} - \vec{r}_o) \frac{d\vec{P}(t)}{dt}$ plays the role of an external polarization current density. Spontaneous emission dynamics of a quantum emitter located in the vicinity of any arbitrary photonic structure can be then obtain by numerically integrating Eqs.(3) or (4) using finite difference time domain method (FDTD) [23] under the initial conditions of zero fields ($\vec{E}(\vec{r}, t=0) = 0$ and $\vec{H}(\vec{r}, t=0) = 0$) and one electronic excitation ($\vec{f}(0) = 1$ or $\vec{P}(0) = 1$). Note that throughout the numerical integration of Eqs. (3) and (4), the delta function $\delta_D(\vec{r} - \vec{r}_o)$ is averaged over one Yee cell [23] which regularize the otherwise singular behavior of the EM Green's function. As we will show later, both Eqs.(3) and (4) give nearly identical results for $|\vec{f}(t)|$ and $|\vec{P}(t)|$.

We now illustrate the power of our proposed technique by computing spontaneous emission dynamics for several different geometries. In all our simulations we consider quantum emitters having transition frequency that corresponds to a free space wavelength $\lambda_o = 1.5 \, \mu m$ and effective dipole moment transition matrix elements of $d_{eg} = -1.342 \times 10^{-28}$ C m [24]. In addition, without any loss of generality we take $L = 4.652$ nm and $A = L^2$ in our idealized 2D and 1D scenarios [24, 25]. Finally we note that in integrating Eq. (4), we assume that the dipoles are oriented perpendicular to the direction of EM wave propagation.

By keeping the above discussion in mind, we first explore the weak coupling regime and we start with two simple examples of 1D and 2D quantum emitters in free space. Under these conditions, the SE life times are found to be $\tau_{1D} = \frac{\varepsilon_0 \hbar c A}{d_{eg}^2 \omega_0} = 0.267$ $ps$ and

$\tau_{2D} = \frac{2\varepsilon_0 \hbar c^2 L}{d_{eg}^2 \omega_0^2} = 27.45$ $ps$, respectively [1, 25]. By integrating Eq. (4) using FDTD under the appropriate initial conditions for the above geometries, we find an excellent agreement between $\left|\vec{P}(t)\right|^2$ and the analytical predictions.

Our next example within the weak-coupling regime is a QE located at the center of a dielectric cylinder of radius 3 $mm$ and relative permittivity $\varepsilon_r = 11.56$ as shown schematically in Fig.3 (a). Figure 3 (b) depicts a comparison between $\left|\vec{P}(t)\right|^2$ as obtained from Eq. (4) (blue solid line) and the predictions of ref. [13] (red circles) for this geometry where an agreement within an error of 1.5% is obtained.

Next we investigate the process of spontaneous emission in the strong coupling regime. Here the SE dynamics are characterized by a non-negligible feedback between the quantum emitter and the surrounding photonic environment. As a result the Markovian approximation breaks down and Wigner-Weisskopf approach fails [11]. Note that in this case, even the numerical scheme of [13] cannot be used to predict non-Markovian effects.

Our first example here is a QE located at the center of a 1D optical cavity that consists of two perfect mirrors. The cavity length $l_x = \frac{\lambda_o}{2}$ is chosen to produce resonant interaction between the fundamental optical mode and the transition frequency of the QE. In this case, Jaynes-Cummings (JC) Hamiltonian [1, 2] predicts an indefinite exchange of excitation between the quantum emitter and the photonic mode (vacuum Rabi splitting) at a frequency $f_R = \sqrt{\frac{\omega_o d_{eg}^2}{\hbar \varepsilon_o l_x A}}$. Our analysis in this simple case indicates a perfect agreement between Eq.(4) and the JC model. A slight variation of the above problem reveals the strength of our technique. In particular, we consider an optical cavity that exhibit degeneracies. In this case, the atomic transition frequency can be in resonance with more than one discrete eigenmode. Before the JC can be used, all these degenerate modes must be first calculated. On the other hand, our FDTD-based method treats all the eigenmodes collectively in one single run. To illustrate this point, we consider a QE located at the center a perfect 2D square cavity of length $l$ as shown schematically in Fig.4 (a). The optical eigenfrequencies of this cavity are given by $\omega_{n,m} = \frac{\pi c}{l}\sqrt{n^2 + m^2}$, where $m$ and $n$ are non-negative integers that represent the mode indices. If the cavity length is taken to be $l = l_1 \equiv \frac{\lambda_o}{\sqrt{2}}$, the QE interacts resonantly only with the fundamental mode of the optical cavity $\omega_{1,1}$ and the situation is similar to the 1D case described above. On the other hand, if we set $l = l_2 \equiv \sqrt{\frac{5}{2}}\lambda_o$, the transition frequency becomes in resonance with two different degenerate eigenmodes associated with $\omega_{1,3}$ and $\omega_{3,1}$ and the JC model predicts Rabi oscillations at frequency $f_R = \sqrt{\frac{2\omega_{1,3}d_{eg}^2}{\hbar\varepsilon_0 l_2^2 L}}$ . Figure 4 (b) depicts the Rabi oscillations dynamics of $\left|\vec{P}(t)\right|^2$ for the two different choices of $l_{1,2}$ (blue/red solid lines). Results obtained from JC Hamiltonian are also presented in the same figure (blue circles and red triangles, respectively) where excellent agreement with Eq. (4) is observed.

In order to further demonstrate the strength of our method, we consider situations where a transition between weak and strong interactions takes place as the physical parameters are varied. In particular, we investigate the case of a QE layer located at the center of a 1D photonic crystal cavity as shown schematically in Fig.5 (a). If the dielectric photonic crystal mirrors are perfect, the QE will resonantly interact with one discrete defect state and SE can be thus described by JC Hamiltonian as before. However, if the Bragg mirrors are made of finite layers, the analysis becomes complicated due to optical coupling to the continuum. As we have mentioned before, techniques based on quasi-normal modes (QNM) [17] and constant flux (CF) states [18] have been proposed to overcome this problem. However, as we noted, computing QNM or CF states is not a straightforward task. Our technique on the other hand circumvents all these difficulties in a straightforward fashion through the FDTD procedure where the effects associated with the radiation modes is simply taken into account by introducing absorbing boundary conditions (ABC) at the edges of the structure. This can be done by either using perfectly matched layer (PML) or by using Mur's boundary conditions in 1D [23]. For illustration purpose, we assume that the photonic band gap (PBG) mirrors shown in Fig.5 (a) consist of quarter wavelength alternating layers of *AlAs* and *GaAs* having refractive indices $n_1 = 2.89$ and $n_2 = 3.37$, respectively. The cavity is assumed to be made of an *AlAs* layer and having a thickness $\lambda_0 / n_2$. Under these conditions, the optical cavity exhibits a localized leaky eigenmode at $\lambda_0$ in addition to radiation modes [26]. When the number of unit cells in each Bragg mirror *N* is very large (ideally infinite), the photonic band gap (PBG) resonator becomes perfect and the situation resembles that of perfect metallic cavity. On the other hand, when *N* is small, the SE process is described by the Purcell effect. However, when the value of *N* lies between these two extremes, an interesting situation arises as the interaction can be affected by the EPs branch cut singularities [15]; thus leading to power-law algebraic decay dynamics [27]. This transition between different coupling regimes is illustrated in Fig.5 (b) where $\left|\vec{P}(t)\right|^2$ is plotted for three different mirrors parameters: $N = 5$ (red dashed line), 10 (blue double dashed line) and 25 (black solid line), respectively.

Finally we note that all these results were repeated by using Eq. (3) instead of the simplified version of Eq. (4) and excellent agreement were also obtained.

In conclusion, we have shown that the process of spontaneous emission from a quantum emitter can be mapped onto a time domain complex oscillator-field model. By utilizing this equivalency, we have presented a unified numerical treatment based on FDTD for spontaneous emission dynamics in both weak and strong coupling regimes as well as their boundary. Finally we have demonstrated the strength of our technique by considering several different examples of quantum emitters located inside different one and two dimensional photonic environments. Our work provides powerful computational tool for engineering integrated quantum devices such as single photon sources.

**Figure captions**

**Figure 1:** (Color online) (a) A schematic of two-level quantum system. (b) Due to negligible off-resonant contribution, spontaneous emission dynamics of such a two-level emitter becomes equivalent to that of a three-level system when the latter is initially excited to the first excited state in the absence of any photonic excitations.

**Figure 2:** (Color online) Ideal three, two and one-dimensional quantum emitters (QEs) are depicted from left to right. The direction of EM wave propagation is indicated by the arrows.

**Figure 3:** (Color online) (a) Top view of a QE located at the center of a dielectric microdisk of radius $3\mu m$ and $\varepsilon_r = 11.56$. (b) Shows the comparison between the predictions of Eq.(4) and the method presented in ref. [13] for the geometry shown in (a).

**Figure 4:** (Color online) (a) A schematic (not to scale) of QE located at the center of a two dimensional perfect metallic cavity. (b) Depicts the vacuum Rabi oscillations associated with the structure in (a) when $l = l_1 \circ \frac{l_o}{\sqrt{2}}$ (blue line) and $l = l_2 \equiv \sqrt{\frac{5}{2}}\lambda_o$ (red line). In the first case the QE is in resonance with the fundamental $\omega_{1,1}$ eigenmode only while in the latter situation it interacts resonantly with the two degenerate eigenmodes $\omega_{1,3}$ and $\omega_{3,1}$. Comparisons with results obtained by using the JC Hamiltonian are also presented in the same figure by using blue circles for $l = l_1$ and red triangles when $l = l_2$.

**Figure 5:** (Color online) (a) A schematic (not to scale) of a QE located at the center of a one dimensional $\lambda_o$ Bragg cavity. The cavity mirrors are made of alternate layers of *AlAs* and *GaAs* having refractive indices $n_1 = 2.89$ and $n_2 = 3.37$, respectively. (b) Rabi Oscillation when the Bragg mirror on each side of the cavity has the following number of unit cells $N = 25$ (Black solid line), 10 (Blue dashed-dotted line) and 5 (Red dashed line). These choices exemplify a transition from strong to weak coupling regimes where the SE dynamics can be directly captured by our method in all regimes.

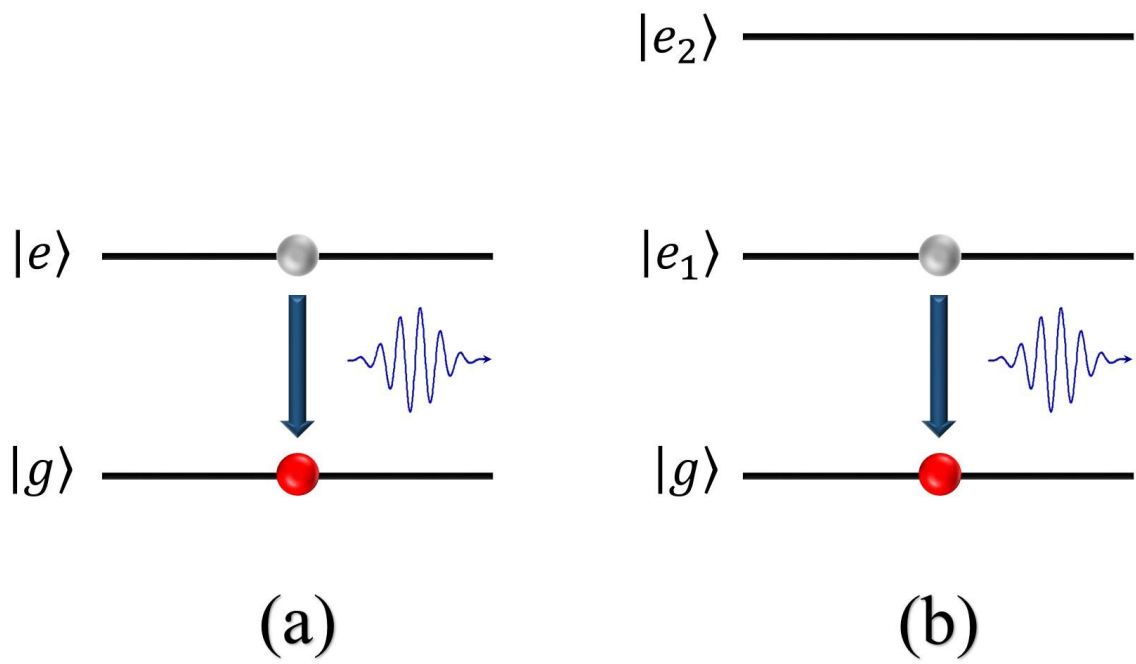

**Fig.1**

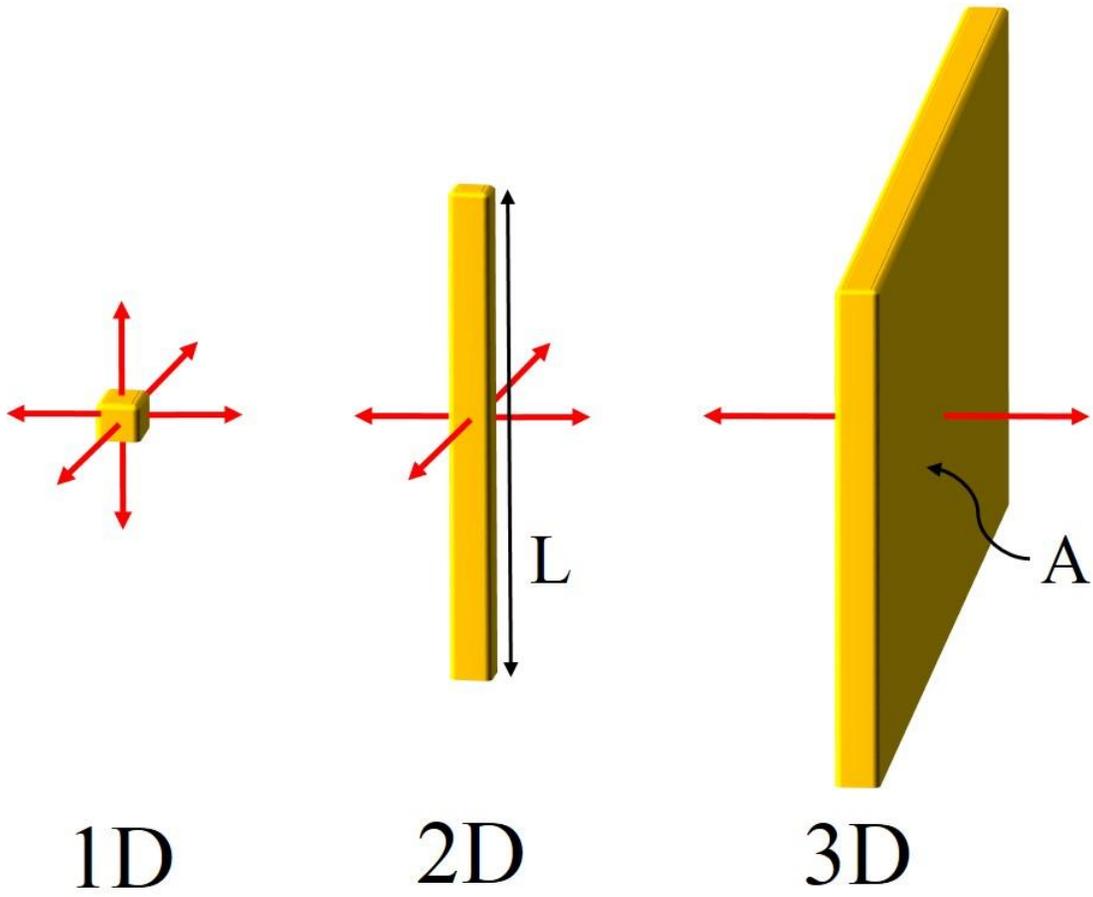

**Fig.2**

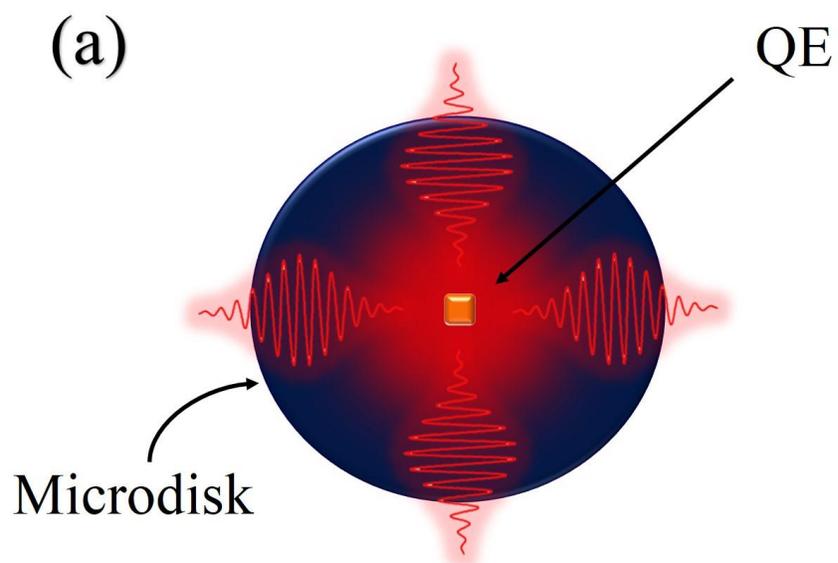

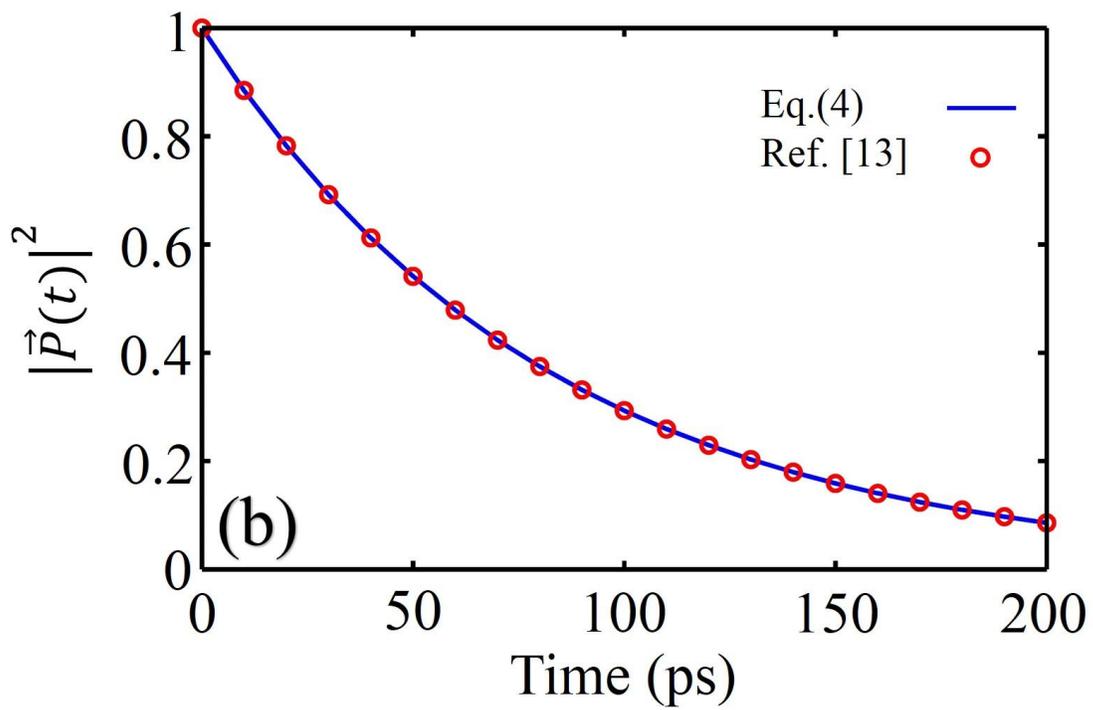

**Fig.3**

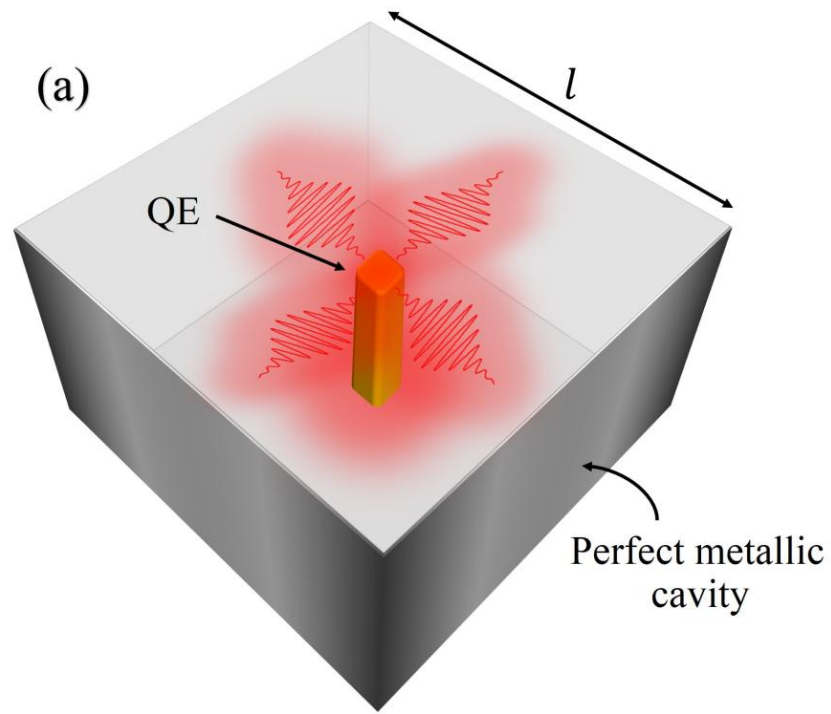
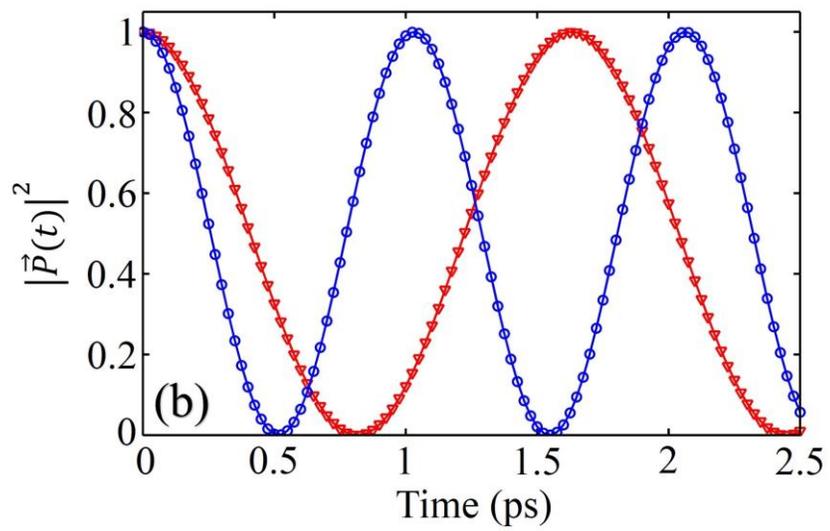

**Fig.4**

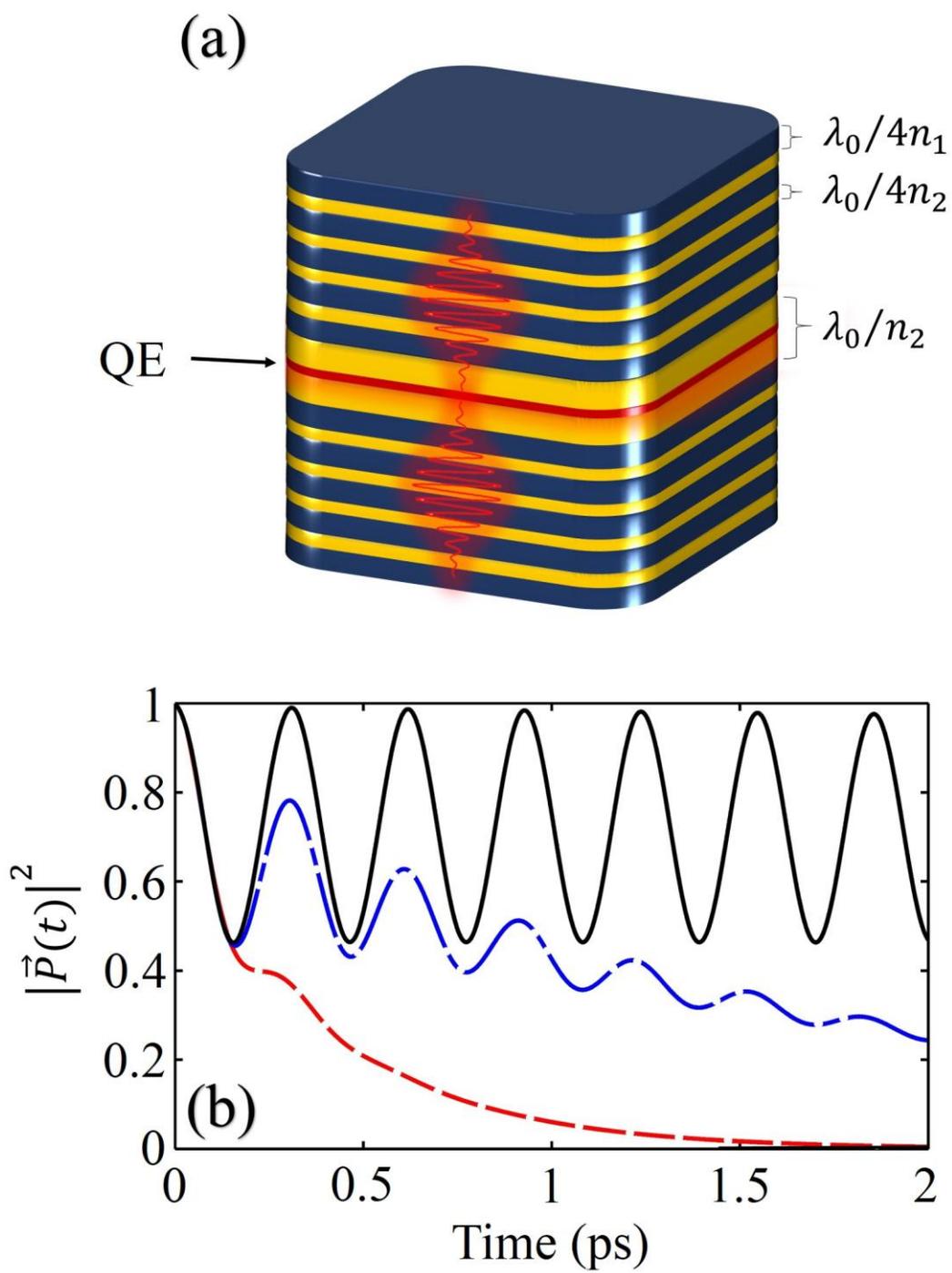

**Fig.5**